\providecommand{\arxivversion}{1}
\documentclass[webpdf,contemporary,large,namedate]{oup-authoring-template}

\newcommand{\societylogo}{}

\ifnum\arxivversion=1
\makeatletter
\def\ps@opening{\let\@oddhead\@empty\let\@evenhead\@empty
  \def\@oddfoot{\hfil\sffamily\fontsize{8bp}{10bp}\selectfont\thepage\hfil}\let\@evenfoot\@oddfoot}
\let\ps@headings\ps@opening
\makeatother
\fi

\usepackage[T1]{fontenc}
\usepackage{mathtools}
\usepackage{bm}
\usetikzlibrary{decorations.pathreplacing, arrows.meta, calc}

\definecolor{metablue}{HTML}{800020}
\definecolor{metafg}{HTML}{1C2B33}
\definecolor{metabg}{HTML}{F1F4F7}
\definecolor{metagray}{RGB}{101,103,107}
\definecolor{theorembg}{HTML}{F2E6E4}
\definecolor{quotebg}{HTML}{D6E4F0}
\definecolor{keyparbg}{HTML}{FAF0F0}   %
\definecolor{keyparhead}{HTML}{5C0016} %

\makeatletter
\renewcommand\paragraph{\@startsection{paragraph}{4}{\z@}%
  {1.5ex \@plus .5ex \@minus .2ex}{-1em}{\normalfont\normalsize\bfseries}}
\makeatother

\theoremstyle{thmstyleone}

\theoremstyle{thmstyletwo}

\theoremstyle{thmstylethree}
\newtheorem{definition}{Definition}

\newtheorem{hypothesis}{Hypothesis}
\newtheorem{model}{Model}

\DeclareMathOperator{\lum}{\lambda}

\DeclareMathOperator{\integ}{\iota}

\newcommand{\bA}{\mathbf{A}}

\newcommand{\bG}{\mathbf{G}}

\newcommand{\bJ}{\mathbf{J}}

\newcommand{\be}{\mathbf{e}}

\newcommand{\bh}{\mathbf{h}}
\newcommand{\bk}{\mathbf{k}}

\newcommand{\br}{\mathbf{r}}

\newcommand{\bu}{\mathbf{u}}

\newcommand{\bx}{\mathbf{x}}
\newcommand{\by}{\mathbf{y}}

\newcommand{\bxi}{\boldsymbol{\xi}}

\newcommand{\cS}{\mathcal{S}}

\newcommand{\cW}{\mathcal{W}}

\newcommand{\pending}[1]{\textcolor{red}{[#1]}}

\newcounter{expt}
\newcommand{\experiment}[2]{\refstepcounter{expt}\label{#1}\paragraph{Experiment~\theexpt: #2.}}
\newcounter{suppc}\renewcommand{\thesuppc}{S\arabic{suppc}}
\newcommand{\suppexp}[2]{\refstepcounter{suppc}\label{#1}\paragraph{\thesuppc: #2.}}

\begin{document}

\journaltitle{Neuroscience of Consciousness}
\DOI{DOI added during production}
\copyrightyear{YEAR}
\pubyear{YEAR}
\vol{XX}
\issue{x}
\access{Published: Date added during production}
\appnotes{Research Article}

\firstpage{1}

\title[Valence as the rate of change of Jacobian scale]{Affective Valence as the Interoceptive\\ Rate of Change of Jacobian Scale}

\ifnum\arxivversion=1
\author[$\ast$]{David Balduzzi}
\corresp[$\ast$]{\href{mailto:dbalduzzi@gmail.com}{dbalduzzi@gmail.com}}
\else
\author[1,$\ast$]{David Balduzzi}
\address[1]{\orgname{\pending{affiliation}}}
\corresp[$\ast$]{Corresponding author. \href{mailto:dbalduzzi@gmail.com}{dbalduzzi@gmail.com}}
\fi

\received{Date}{0}{Year}
\revised{Date}{0}{Year}
\accepted{Date}{0}{Year}

\abstract{Affect is widely held to depend on interoception. This paper proposes that valence is the rate at which interoception moves a self-sustaining system's responsiveness relative to its terminal set (roughly, death). Responsiveness is formalised as the scale of the system's Jacobian and interoception as its internal inputs: receptors, signals and forecasts of its energy reserve. 
Changes in responsiveness that bypass interoception, such as brain damage, are not felt. Language models have no valence because they are not self-sustaining. The chain rule splits valence over the system's parts and the causes that affect its internal inputs. 
With the log scale, these splits account for need-dependent pleasure, diffuse hunger and localised pain, negativity bias, and relief when a threat fails to arrive. 
In a minimal model of a body:
  (1) Blocking the reserve signal leaves depletion unfelt, and driving it produces hunger with an undepleted reserve.
  (2) Removing receptors or their gain, or driving receptors, reproduces congenital insensitivity to pain, asymbolia and phantom pain.
  (3) A receptor gain learned to sustain the body signals local damage to the whole body, so an injury hurts in proportion to its cost.
  (4) If the reserve also modulates the gain, the body learns to suppress pain due to receptors as energy runs low. 
  (5) When it learns to maximise its own scale, it becomes a wirehead.
The account explains when changes are felt, where they are felt, and when feelings mislead.
}


\maketitle

\section{Introduction}
\label{sec:intro}

A meal tastes better when you are hungry than sated~\citep{cabanac1971}. The stimulus is the same; your bodily needs have changed. Feeling good or bad, valence, largely belongs to the body. Yet not everything that happens to the body is felt, not even damage. Brain tissue has no nociceptors~\citep{raywolff1940}: a stroke can destroy a large region of cortex without pain, although it can change, sometimes profoundly, what the patient experiences and does. A small lesion on the pathways that carry signals from the body can cause severe pain~\citep{klit2009}, or leave pain sensed but no longer unpleasant~\citep{berthier1988}. Damage is not felt unless it is signalled. Feeling and harm come apart in other ways: an amputee can feel severe pain in an absent limb~\citep{flor2006}, while children born without nociception injure themselves badly without being hurt~\citep{nagasako2003}. Feelings can have locations: the pain from a cut finger is more localised than hunger.

There is broad agreement that valence is rooted in interoception and homeostatic regulation~\citep{craig2002,damasio2013,seth2013,barrett2015}, and several accounts identify it with a \emph{rate}: of progress toward goals~\citep{carver1990}, of change in free energy or prediction error~\citep{joffily2013,hesp2021,vandecruys2017}, of reduction of a homeostatic deficit~\citep{keramati2014}. 

These accounts leave four questions open. What fixes the reference point, other than a goal or set point chosen by the modeller? Why are some losses felt and others, like brain damage, not? What determines where a feeling is located? When and why does feeling misrepresent what is happening to the system?

\subsection{The hypothesis}
\label{ssec:hyp}

This paper answers all four questions with one quantity. In plain terms, \emph{feeling good is registering, through one's own internal signals, that one is moving away from one's terminal set (roughly, death); feeling bad is registering that one is moving toward it; and the faster the motion, the stronger the feeling.}

The terminal set is determined by the first-order (Jacobian) structure of the system. A companion paper~\citep{balduzzi2026g} proposes that the structure of experience is shaped by the system's Jacobians, the sensitivities of its actions to its inputs and parameters, where actions include the firing of neurons. Here, only the overall size or log scale of the Jacobian is needed, $\lum:=\log\|\bJ_\cS\|$, which measures how responsive the system is. The terminal set is where every response has vanished, $\lum=-\infty$. Valence is the rate at which $\lum$ moves, \emph{counting only motion due to the system's internal inputs}: receptors, signals of its energy reserve, and forecasts built on them (Definition~\ref{def:valence}). Motion of the scale that passes through no internal input, such as an anaesthetic agent or unsignalled damage, changes experience but is not felt as a change.

The proposal concerns \emph{self-sustaining} subjects, systems that experience as one and that run down toward their terminal set unless they act (e.g. eating) to pull themselves back: systems with something at stake. The chain rule splits valence exactly over the system's parts, which says \emph{where} a change is felt, Eq.~\eqref{eq:locate}, and over the causes that affect its internal inputs, which says \emph{why} and whether what is felt is grounded or ungrounded, Eq.~\eqref{eq:split}. Together with the logarithmic scale, these two decompositions account for need-dependent pleasure, diffuse hunger and localised pain, negativity bias, dread and relief, and the clinical dissociations of pain, Table~\ref{tab:phen}.

\subsection{Overview}
\label{ssec:overview}

The methods, \S\ref{sec:methods}, defines the terminal set, self-sustaining systems and valence, and states the hypothesis. The results, \S\ref{sec:results}, builds a minimal differentiable model of a body, with an energy reserve, tissue and receptors, and enables its parts one at a time, so each phenomenon can be traced to a single term. The discussion, \S\ref{sec:discussion}, relates the account to existing work, asks which systems are self-sustaining, and describes how valence could be measured in practice.

\paragraph{Contributions.}
\begin{enumerate}
  \item \emph{A hypothesis on valence}: feeling is motion of the scale $\lum$ relative to the system's terminal set (roughly, death), which is fixed by its physics, rather than chosen by the modeller, \S\ref{sec:hypV}.
  \item \emph{An interoception condition}: only motion of $\lum$ due to internal inputs is felt. The condition separates organisms from language models and externally powered machines. In the model, the same gain is unfelt when set from outside and felt when the system computes it from its own reserve, \S\ref{ssec:dial}.
  \item \emph{Two exact decompositions} of valence, over locations, Eq.~\eqref{eq:locate}, and over causes, Eq.~\eqref{eq:split}, which account for the phenomena in Table~\ref{tab:phen}, and a split of the Jacobian's motion into scale, which is valence, and shape, which locates it, \S\ref{sec:hypV}.
  \item \emph{Dissociations in a model of a body}: blocking the signal of the reserve leaves depletion unfelt and driving it produces hunger with an undepleted reserve; removing receptors or their gain reproduces congenital insensitivity and asymbolia, and driving receptors produces phantom pain, \S\S\ref{ssec:reading},~\ref{ssec:pain}.
  \item \emph{Forecasts.} A forecast is an internal input: valence moves from a drain to the cue that predicts it, relief comes when a threat fails to arrive, and the same forecast error is felt more the nearer the system is to $Z$, \S\ref{ssec:allo}.
  \item \emph{Learned feeling}: when the receptor gain is learned to sustain the body, injuries come to hurt in proportion to their cost, and hunger comes to suppress pain due to receptors. Learning to maximise the log scale itself produces a wirehead, \S\ref{ssec:learned}.
  \item \emph{Consequences for organisms}: known findings the account reproduces, and a few it adds, Table~\ref{tab:pred}.
\end{enumerate}

\subsection{Related work}

A substantial literature argues that metabolism and interoception are fundamental to affect~\citep[see][]{damasio2003,thompson2010,lindquist2012,seth2025}. Closest to Hypothesis~\ref{h:valence} are proposals that valence tracks a \emph{rate} rather than a level. In control-process theory, affect reflects the rate of progress toward or away from goals~\citep{carver1990}. In active inference, valence has been identified with the rate of change of free energy~\citep{joffily2013} and with changes in confidence in one's action model~\citep{hesp2021}; on error-dynamics accounts, it tracks the rate at which prediction error is reduced, relative to the expected rate~\citep{vandecruys2017,kiverstein2019}. Homeostatic reinforcement learning defines reward as reduction in distance to a set point~\citep{keramati2014}, and homeostasis has been proposed as a design principle for machines that feel~\citep{man2019}.

The most developed account of affect grounded in self-maintenance is interoceptive inference~\citep{seth2015,barrett2015,seth2016}, which treats the brain as a regulator of the body's essential variables, and identifies feelings with predictions about the causes of interoceptive signals. The paper shares its cybernetic roots~\citep{ashby1956}. Here, as there, affect is interoceptive (\S\ref{sec:hypV}); forecasts are one kind of internal input (\S\ref{ssec:allo}), and feelings that mislead are errors of the machinery that gets things right, misinference there and ungrounded signals here.

The following features distinguish this account. \emph{Valence is one quantity}, the rate of change of the log scale of the system's Jacobian due to its internal inputs; it needs no generative model, and is not progress in a space of goals or beliefs. \emph{The reference point} is the terminal set, fixed by the system's first-order structure rather than extrinsically stipulated. \emph{Valence depends on causal structure}: the same gain is felt when the system computes it from its own variables, and not when it is set from outside (\S\ref{ssec:dial}); the same depletion is felt when signalled, and not when its signal is blocked. \emph{Intensity depends on spread}~\citep{bergson1910}: a signal confined to one part is diluted by that part's share. \emph{The change in scale decomposes exactly}, over locations and over causes, Eqs.~\eqref{eq:locate} and~\eqref{eq:split}, so where valence is felt, and whether it matches what is happening to the essential variables, are computed rather than argued. 

\section{Methods}
\label{sec:methods}

\subsection{Jacobians and measures}
\label{ssec:jacobians}

Consider a system $\cS$ with inputs $\bx\in\mathbb{R}^d$, parameters $\cW\in\mathbb{R}^p$ and actions $\by\in\mathbb{R}^n$, called outputs in~\citet{balduzzi2026g}: anything the system emits, from the movement of a limb to the firing of a neuron. Its \emph{input} and \emph{parameter} Jacobians are the $n\times d$ and $n\times p$ matrices
\begin{equation*}
  \bJ_{\by\leftarrow \bx} = \nabla_{\bx} \by(\bx, \cW)
  \quad\text{and}\quad
  \bJ_{\by\leftarrow \cW} = \nabla_{\cW} \by(\bx, \cW).
\end{equation*}
The input Jacobian describes which aspects of the world the system is differentially sensitive to; the parameter Jacobian describes the sensitivity of the levers it pulls to modify its own behaviour. Inputs and actions can span several layers and time steps.

\begin{definition}
  \label{def:scale}
  Given a matrix $\bA$ with singular values $\sigma_1\ge\dots\ge\sigma_r>0$, its \textbf{scale} is the Frobenius norm $\|\bA\| = \big(\sum_i\sigma_i^2\big)^{1/2} = \big(\sum_{ij}A_{ij}^2\big)^{1/2}$. It is zero iff $\bA\equiv 0$, and $\|\alpha\bA\|=|\alpha|\,\|\bA\|$.
\end{definition}
\noindent
The following hypothesis, connecting physical interactions to phenomenal experience, was proposed and investigated in~\citet{balduzzi2026g}.
\renewcommand{\thehypothesis}{G}
\begin{hypothesis}
  \label{h:first}
  The first-order structure of physical interactions, Jacobians, shapes the structure of phenomenal experience. Experience registers differences to the extent that they make a difference (\textbf{difference}), and extends to the extent that interactions connect (\textbf{connection}).
\end{hypothesis}
\noindent
Two scale-invariant measures, transparency and integrity, were introduced in~\citet{balduzzi2026g}. Transparency is an effective rank: how many independent directions are transmitted. Integrity is a normalised algebraic connectivity of the graph of interactions: how hard the system is to split. Integrity is zero if a system decomposes into noninteracting parts.

The companion paper also considers how to identify \emph{subjects} of experience: where one experiencing system ends and another begins. It provisionally proposes local maxima of integral rank, units, as candidates for subjects. For the purposes of this paper, the important point is that subjects do not decompose into noninteracting parts; integrity must satisfy $\integ(\bJ_\cS)>0$ (\S\ref{sec:hypV}).

\paragraph{Why Jacobians.}
A Jacobian can be estimated by perturbing a system and recording the change, as in receptive-field mapping, stimulation studies and perturbational measures of consciousness~\citep{massimini2005,casali2013}. Working with derivatives has several advantages. They are the language of rate and dynamical models of the brain, where gain modulation is a basic operation~\citep{salinas2000,ferguson2020}. Stability and susceptibility, which fall under anaesthesia~\citep{tagliazucchi2016}, are read from the Jacobian. They only need the current state, not the full repertoire of states. They are state-dependent, like affect, and cheap to compute. Finally, the chain rule makes the decompositions of \S\ref{sec:hypV} exact.

\subsection{The terminal set}
\label{ssec:zero}

To get at affect, a yardstick is needed that determines what is better or worse for the system. Fix a system $\cS$ and a \emph{window}, the span of time steps over which its inputs and actions are collected. Its \emph{state variables} are the quantities it carries or computes within the window, such as its actions, hidden activations, parameters, energy reserves, tissue condition, or receptor activity (but not external inputs). A \emph{state} $s$ of $\cS$ is the configuration at which its Jacobians are evaluated: the inputs and the values of the state variables.
\begin{definition}
  \label{def:zero}
  Let $\bJ_\cS(s) := \bJ_{\by\leftarrow\cW}(s)$ be the parameter Jacobian at $s$. The \textbf{terminal set} of $\cS$ is $Z := \{ s : \|\bJ_\cS(s)\| = 0 \}$.
\end{definition}
\noindent
The terminal set $Z$ is defined on the \emph{parameter} Jacobian because it concerns the system's ability to \emph{act}: it is where nothing the system does makes a difference, because every action has stopped responding. A limb whose tissue is destroyed contributes a zero row to $\bJ_\cS$, but the rest of the system still responds; the system has moved toward $Z$ only by that part's share of the scale (Eq.~\eqref{eq:locate}). Whether that motion is felt depends on whether it affects an internal input: unsignalled damage is not felt, and pain, in which damage affects receptors and through them activity everywhere, is (\S\ref{ssec:pain}). Anaesthesia and deep sleep move the system toward $Z$ without reaching it (\S\ref{sec:hypV}). In the model below $Z$ is absorbing: once the reserve is empty, activity ends.

\begin{figure}[htbp]
  \centering
  \resizebox{\columnwidth}{!}{  \begin{tikzpicture}[x=1cm,y=1cm]

  \path (-0.3,-0.58) rectangle (6.7,3.20);

  \coordinate (O) at (0.55,0.5);

  \def\angA{6}  \def\lenA{2.0}
  \def\angB{20} \def\lenB{2.7}
  \def\angC{46} \def\lenC{2.3}
  \def\angD{62} \def\lenD{1.7}

  \foreach \RA/\RL in {\angA/\lenA, \angB/\lenB, \angC/\lenC, \angD/\lenD}{
    \pgfmathsetmacro{\ca}{cos(\RA)}
    \pgfmathsetmacro{\sa}{sin(\RA)}
    \foreach \i in {1,...,14}{
      \pgfmathsetmacro{\tA}{(\i-1)/14}
      \pgfmathsetmacro{\tB}{\i/14}
      \pgfmathsetmacro{\grayPct}{80-40*(\tA+\tB)}
      \draw[line width=1pt, line cap=round, color=metagray!\grayPct!metablue]
        ($(O)+(\tA*\RL*\ca,\tA*\RL*\sa)$) -- ($(O)+(\tB*\RL*\ca,\tB*\RL*\sa)$);
    }
    \fill[metablue] ($(O)+(\RL*\ca,\RL*\sa)$) circle (1.6pt);
  }

  \pgfmathsetmacro{\cA}{cos(\angA)} \pgfmathsetmacro{\sA}{sin(\angA)}
  \pgfmathsetmacro{\cB}{cos(\angB)} \pgfmathsetmacro{\sB}{sin(\angB)}
  \pgfmathsetmacro{\cC}{cos(\angC)} \pgfmathsetmacro{\sC}{sin(\angC)}
  \pgfmathsetmacro{\cD}{cos(\angD)} \pgfmathsetmacro{\sD}{sin(\angD)}
  \def\tin{0.32}
  \draw[metagray, line width=0.9pt, dash pattern=on 2.2pt off 1.6pt]
    ($(O)+(\tin*\lenA*\cA,\tin*\lenA*\sA)$) --
    ($(O)+(\tin*\lenB*\cB,\tin*\lenB*\sB)$) --
    ($(O)+(\tin*\lenC*\cC,\tin*\lenC*\sC)$) --
    ($(O)+(\tin*\lenD*\cD,\tin*\lenD*\sD)$);

  \draw[metablue, line width=0.7pt, opacity=0.55]
    ($(O)+(\lenA*\cA,\lenA*\sA)$) --
    ($(O)+(\lenB*\cB,\lenB*\sB)$) --
    ($(O)+(\lenC*\cC,\lenC*\sC)$) --
    ($(O)+(\lenD*\cD,\lenD*\sD)$);

  \node[font=\tiny\itshape, text=metagray, anchor=south west]
    at ($(O)+(\lenD*\cD,\lenD*\sD)+(0.35,0.12)$)
    {$\bJ_{y_i\leftarrow \cW}$};

  \node[circle, fill=metagray, draw=white, line width=0.8pt, minimum size=0.20cm, inner sep=0pt] at (O) {};

  \node[font=\footnotesize\itshape, text=metagray, anchor=north, align=center]
    at ($(O)+(0,-0.20)$) {$m\rightarrow0$:\\bundle shrinks};
  \node[font=\footnotesize\itshape, text=metablue!80!black, anchor=west, align=center]
    at ($(O)+(\lenA*\cA,\lenA*\sA)+(0.30,-0.05)$) {$m\rightarrow1$:\\bundle expands};

  \end{tikzpicture}}
  \caption{\textbf{Jacobian space in polar coordinates.} Each ray is a Jacobian: shape and scale are direction and length; the origin is the terminal set $Z$. Here the rays are the rows $\bJ_{y_i\leftarrow\cW}$ of $\bJ_\cS$, all scaled by a metabolic gain $m$ (\S\ref{ssec:dial}): as $m\to0$ the bundle collapses toward the origin, the terminal set $Z$; as $m\to1$ it extends to its full length. Scale changes; shape does not.}
  \label{fig:affect}
  \figalttext[Jacobian space drawn in polar coordinates, with a bundle of rays that shrinks or expands with a metabolic gain.]{Rays from an origin, each a row of the Jacobian, with direction showing shape and length showing scale. As the gain goes to zero the bundle collapses toward the origin, the terminal set; as it goes to one the bundle fans out. Scale changes while shape does not.}%
\end{figure}
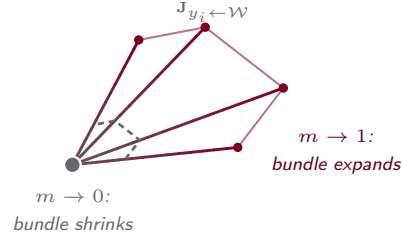

The terminal set is the \emph{origin} of \emph{Jacobian space}, the space which, by Hypothesis~\ref{h:first}, shapes experience. In Jacobian space $Z$ is a single point; in the system's state space it is a large set: intuitively, all states in which the system can no longer respond (e.g. because it is dead).

Jacobian space factorizes into polar coordinates, Fig.~\ref{fig:affect}. Let $\bJ_\cS=\|\bJ_\cS\|\,\hat\bJ_\cS$ where $\|\hat\bJ_\cS\|=1$. Factors $\|\bJ_\cS\|$ and $\hat\bJ_\cS$ are the \emph{scale} and \emph{shape}. Transparency and integrity are scale-invariant, so only depend on the shape. The scale is one-dimensional with $Z$ its unique boundary point. 

Write $\lum:=\log\|\bJ_\cS\|$ for the \textbf{log scale}: $Z$ lies at $\lum=-\infty$. Within a window, $\bJ_{\cS,t}$ collects the rows of $\bJ_\cS$ belonging to actions at time $t$, and $\lum_t:=\log\|\bJ_{\cS,t}\|$ is the log scale at time $t$.

\subsection{Self-sustaining systems}
\label{ssec:selfsust}

The terminal set \emph{$Z$ is attracting:} responsive systems are energetically expensive, and a target for parasites, predators and the second law~\citep{jonas1966}. Without effort, trajectories drift toward $Z$.

\begin{definition}
  \label{def:loop}
  The \textbf{essential variables} of $\cS$ are the state variables $\be$ that (i) its actions affect, (ii) affect the scale, and (iii) are \textbf{precarious}, can potentially move $\cS$ toward $Z$:
  \begin{equation*}
    \bJ_{\be\leftarrow\by}\neq0,\qquad \frac{d\lum}{d\be}\neq0,\qquad \min_{\by}\Big\langle \frac{d\lum}{d\be},\,\dot\be(\by)\Big\rangle<0,
  \end{equation*}
  where $\frac{d\lum}{d\be}$ is the total derivative, the effect of $\be$ on the scale directly and indirectly through state variables, and the minimum is over all available actions. $\cS$ is \textbf{self-sustaining} if it has essential variables. Its \textbf{internal inputs} are the state variables $\bu$ with $\bJ_{\bu\leftarrow \be}\neq0$ and $\partial_{\bu}\lum\neq0$ that do not run down on their own.
\end{definition}
\noindent
Energy reserves and tissue condition are essential variables. Exposure is not enough: a language model's weights can be damaged, but it does not avoid or repair the damage, so they are not essential variables. Essential variables can run down in multiple directions: temperature going to either extreme moves the system toward $Z$. Self-sustaining asks what the system \emph{can} do, not what it is currently doing: an animal that is asleep, idle or restrained remains self-sustaining. 

\begin{figure}[t]
  \centering
  \resizebox{\columnwidth}{!}{\begin{tikzpicture}[x=1cm,y=1cm, font=\small, line cap=round, line join=round,
  lab/.style={font=\scriptsize, inner sep=1pt, align=center},
  ar/.style={-{Stealth[length=1.8mm]}, thick, metafg!70},
  felt/.style={-{Stealth[length=2mm]}, very thick, sigC},
  dir/.style={-{Stealth[length=1.8mm]}, thick, metagray, dashed},
  fal/.style={-{Stealth[length=1.8mm]}, thick, falC}]
\definecolor{essC}{HTML}{C08A1E}\definecolor{sigC}{HTML}{2C6FA8}\definecolor{falC}{HTML}{B03030}
\draw[essC, line width=1pt, rounded corners=3pt, fill=white] (-0.4,0.75) rectangle (0.4,1.75);
\fill[essC!50, rounded corners=2pt] (-0.35,0.8) rectangle (0.35,1.3);
\node[lab, text=essC!70!black, anchor=north] at (0,0.62) {essential\\[-1pt]variable $\be$};
\node[lab, text=metagray, anchor=north] at (0,0.12) {$E$, $h_i$};
\draw[ar] (0,2.55) node[lab, above]{actions $\by$} -- (0,1.85);

\draw[ar, essC!70!black] (-0.45,1.0) to[out=180,in=90] node[lab, left, pos=0.55]{runs\\[-1pt]down} (-1.0,-0.35);
\fill[metafg] (-1.0,-0.5) circle (2.2pt) node[lab, below=2pt]{$Z$};
\begin{scope}[shift={(2.5,-0.05)}]
\draw[sigC, line width=1pt, fill=sigC!8] (-0.48,0) arc[start angle=180, end angle=0, radius=0.48] -- cycle;
\foreach \t in {20,50,90,130,160} \draw[sigC] (\t:0.36) -- (\t:0.46);
\draw[sigC, very thick] (0,0) -- (55:0.4); \fill[sigC] (0,0) circle (1.6pt);
\node[lab, text=sigC, anchor=north] at (0,-0.1) {internal\\[-1pt]input $\bu$};
\node[lab, text=metagray, anchor=north] at (0,-0.6) {$\hat E$, $r_i$};
\coordinate (U) at (0,0.5); \coordinate (Uw) at (-0.5,0.15); \coordinate (Ue) at (0.5,0.15);
\end{scope}
\begin{scope}[shift={(4.6,0.75)}]
\foreach \ang/\len in {8/1.0, 24/1.25, 42/1.1, 60/0.85}
  \draw[metablue!80, line width=1.1pt] (0,0) -- (\ang:\len) node[circle, fill=metablue, inner sep=0.9pt]{};
\fill[metafg] (0,0) circle (1.6pt);
\node[lab, anchor=north] at (0.55,-0.1) {log scale $\lum$};
\coordinate (L) at (0.0,0.15);
\end{scope}
\draw[dir] (0.45,1.45) to[out=10,in=150] node[lab, above, text=metagray]{direct: not felt} (4.45,1.25);
\draw[felt] (0.45,1.0) to[out=-20,in=170] node[lab, above right=-1pt, text=sigC, pos=0.5]{senses} (Uw);
\draw[felt] (Ue) to[out=10,in=-150] node[lab, below right=-2pt, text=sigC, pos=0.55]{felt: $v$} (L);
\draw[fal] (2.5,-1.55) node[lab, below, text=falC]{other drive $\bxi$: phantom, false cue, drug} -- node[lab, right, text=falC]{ungrounded} (2.5,-1.0);
\end{tikzpicture}}
  \caption{\textbf{Essential variables and internal inputs.} Actions (e.g. eating) pull essential variables back from $Z$; left alone, they run down. They can affect the log scale directly, which is not felt, or through internal inputs, which is valence. Other drives on internal inputs are ungrounded, Eq.~\eqref{eq:split}.}
  \label{fig:routes}
  \figalttext[Diagram of how essential variables reach the log scale, directly or through internal inputs.]{Actions pull an essential variable back from the terminal set; left alone it runs down. The essential variable reaches the log scale directly, which is not felt, or through an internal input, which is valence. A separate drive on the internal input from anything else is ungrounded.}%
\end{figure}
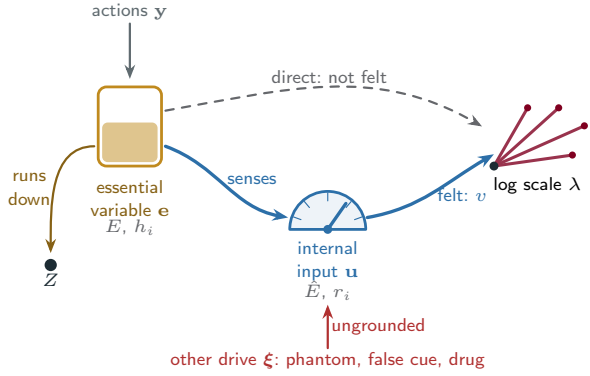

Internal inputs signal what is at stake: receptor activity, a hormone level, a forecast. Essential variables can reach the scale by two routes, directly and through internal inputs, Fig.~\ref{fig:routes}; only the second is felt (\S\ref{sec:hypV}). Whether a variable is an internal input depends on the subject's boundary.

Definition~\ref{def:loop} reformulates Jonas's needful freedom~\citep{jonas1966} and Di Paolo's precariousness~\citep{dipaolo2005,dipaolo2009} in terms of the terminal set. Essential variables are taken from Ashby~\citep{ashby1956}, but here they are picked out by the system's Jacobians and dynamics rather than stipulated.

\subsection{Valence}
\label{sec:hypV}

A structure of differential sensitivities has no valence; Jacobians are neither for nor against anything. They have no stake in themselves.

\renewcommand{\thehypothesis}{V}
\begin{hypothesis}
  \label{h:valence}
  Valence arises only in self-sustaining subjects, where motion toward or away from $Z$ is felt as negative or positive to the extent that it is due to their internal inputs.
\end{hypothesis}
\noindent
Valence is interoceptive: it depends on internal signals such as nociception, signals of energy and the forecasts built on them~\citep{craig2002}. Valence only arises in self-sustaining systems because internal inputs and essential variables only exist there (Definition~\ref{def:loop}). By Hypothesis~\ref{h:first}, all of the system's activity shapes its experience, but only the part of a rescaling due to interoception is felt: valence is the experience, through interoception, of a rescaling of experience.

\begin{definition}
  \label{def:valence}
  Let $\bu$ denote the internal inputs of $\cS$. The \textbf{valence} of $\cS$ is the change of the Jacobian's \emph{log scale} due to its internal inputs,
  \begin{equation*}
    v = \big\langle \partial_{\bu}\lum,\ \dot\bu\big\rangle
    = \big\langle \partial_{\bu}\lum,\ \tfrac{d\bu}{dt}\big\rangle ,
  \end{equation*}
  where each partial derivative holds the essential variables and the other internal inputs fixed. $\lum_t$ is taken over the rows of $\bJ_\cS$ at time $t$.
\end{definition}
\noindent
Since essential variables and internal inputs are not parameters, it is worth mentioning how they appear in the parameter Jacobian $\bJ_\cS$. In Model~\ref{m:full} they act as gains on activity, so each row of $\bJ_\cS$ is a product of gains and a row set by the network. The derivative $\partial_\bu\lum = \partial_\bu\log \|\bJ_{\by\leftarrow\cW}(s)\|$ then says how the scale of the first-order structure changes as the state moves, which makes valence a second-order quantity.

Valence belongs to the system, and not just to the stimulus: the same stimulus can be felt differently. Parameters are also state variables: when learning modifies them the change is felt iff parameters are internal inputs on the timescale of learning. The rest of $\dot\lum$ is not felt: motion of essential variables acting \emph{directly}, as when damaged tissue dims its own activity, and of any activity that no essential variable affects.

The logarithm has three virtues. It is unit-free: changing units multiplies $\|\bJ_\cS\|$ by a constant, which the derivative of the logarithm removes. It places $Z$ at $-\infty$, so the same absolute loss matters more the closer the system is to its end. Finally, it separates a gain from what it scales: if $\bJ_\cS=m\bJ_0$, then $\lum=\log m+\log\|\bJ_0\|$. That relative rather than absolute change is felt is the form of the Weber--Fechner law~\citep{fechner1860}.

Motion toward the terminal set is negative because it is \emph{worse for the system}: $Z$ is absorbing (hitting $Z$ ends the system), and drift carries the system toward $Z$ by default, so approach is a failure of its effort to persist. A system holding steady has $v\approx0$ wherever it is; a system close to $Z$ and retreating from it feels relief.

\paragraph{Why interoception: anaesthesia and brain damage.}
Under general anaesthesia, responsiveness moves toward $Z$ further and faster than in almost any other circumstance. However, anaesthesia does not feel terrible, so motion toward $Z$ is not always felt.

Anaesthesia removes two things. It breaks down cortical effective connectivity~\citep{massimini2005,ferrarelli2010} and it silences the internal inputs: an anaesthetised patient cannot be roused by damage that would wake a sleeper~\citep{brown2010}, so nociceptors still fire but their signal no longer changes what the brain does, and the patient's essential variables are externally sustained by a ventilator, fluids and warming. Anaesthesia lowers responsiveness directly and is signalled by nothing.

To feel valence, a system must be a subject, and motion of its scale must be due to interoception. An aware patient under neuromuscular block cannot move but can suffer acutely~\citep{pandit2014}, since the block lies outside the experiencing system and leaves the internal inputs intact. A sleeper can be aroused, so interoception is idle, not cut. Under anaesthesia interoception fails, and the subject itself may fragment as effective connectivity breaks down.

Brain damage can go unfelt in a conscious subject. A patient with a focal lesion is conscious and their interoception is intact, but brain tissue has no nociceptors~\citep{raywolff1940}: losing it lowers responsiveness directly, through no internal input, and is not felt. A lesion is felt only when it falls on the pathways that carry internal inputs. Driving them gives central post-stroke pain~\citep{klit2009}; cutting them gives pain asymbolia~\citep{berthier1988}. 

More abstractly, only a signal can (i) be mistaken and (ii) guide action: a regulator must carry information about what it regulates~\citep{conant1970}. Damage to direct routes degrades the system without informing it.

\paragraph{Valence decomposed over locations.} Write $\bJ_i$ for the row of $\bJ_\cS$ belonging to action $i$ and $\ell_i=\|\bJ_i\|$ for its norm, so that $\|\bJ_\cS\|^2=\sum_i \ell_i^2$. Valence splits exactly over actions, $v=\sum_i v_i$, where, with $\tfrac{d}{dt}$ taken through the internal inputs as in Definition~\ref{def:valence},
\begin{equation}
  \label{eq:locate}
  v_i=\frac{\langle\bJ_i,\dot\bJ_i\rangle}{\|\bJ_\cS\|^2}=w_i\,\tfrac{d}{dt}\log \ell_i,\qquad
  w_i=\frac{\ell_i^2}{\|\bJ_\cS\|^2},
\end{equation}
so each action's change counts in proportion to its share of the scale. A change confined to one action is diluted by that action's share; a change across all actions is not. This is why a subject must not decompose into noninteracting parts: otherwise the shares would weigh one part's change against a part it never interacts with, diluting one subject's pain by another's activity.

The \textbf{residual field} expresses valence's spatial structure:
\begin{equation*}
  \delta_i := v_i-w_i\,v.
\end{equation*}
It sums to zero, vanishes when the change is a pure rescaling, and is concentrated on the actions where motion occurs. Valence is \emph{localised} to the extent that $\delta$ is concentrated. For a scalar gain $m$ set by an internal input, $\bJ_\cS=m\bJ_0$, so $v=\tfrac{d}{dt}\log m$ and $\delta=0$: valence is diffuse. For a gain $g_i$ on each action, set by an internal input, $v=\sum_i w_i\,\tfrac{d}{dt}\log g_i$, so a gain that changes at one location more than the rest localises valence.

Location and intensity are distinct: a change to one action is maximally localised, but faint because it is diluted. Terms of opposite sign can coexist: $v_i$ can be positive at some locations and negative at others, and the grounded and ungrounded terms of Eq.~\eqref{eq:split} can oppose, so mixed feelings are concurrent terms of opposite sign rather than a midpoint~\citep{larsen2001}.

\paragraph{Geometry.}
The Jacobian moves through the internal inputs at velocity $\dot\bJ_\cS = \sum_k \partial_{u_k}\bJ_\cS\,\dot u_k$, a mixed second derivative of the actions in parameters and internal inputs. Decompose as $\dot\bJ_\cS=v\,\bJ_\cS+\dot\bJ_\perp$, with $\langle\bJ_\cS,\dot\bJ_\perp\rangle=0$. The first part is radial motion, Fig.~\ref{fig:affect}; its coefficient is valence, $v=\langle\bJ_\cS,\dot\bJ_\cS\rangle/\|\bJ_\cS\|^2$. The second is tangential; it changes the shape $\hat\bJ_\cS$ and adds nothing to valence. Projecting $\dot\bJ_\perp$ onto each row of $\bJ_\cS$ gives the residual field, $\delta_i=\langle\bJ_i,(\dot\bJ_\perp)_i\rangle/\|\bJ_\cS\|^2$, which sums to zero because $\dot\bJ_\perp$ is orthogonal to $\bJ_\cS$. A feeling is located by the part of the shape change that shifts size between actions. A scalar gain has no shape part, so it is diffuse.

\paragraph{Valence decomposed over causes.}
Dividing what affects the internal inputs $\bu$ into essential variables $\be$ (reserves, tissue) and everything else $\bxi$ (phantom drives, forecast errors, drugs, cues that predict nothing), $\dot\bu=\bJ_{\bu\leftarrow \be}\dot\be+\bJ_{\bu\leftarrow\bxi}\dot\bxi$, yields
\begin{equation}
  \label{eq:split}
  v = \underbrace{\langle\partial_\bu\lum,\,\bJ_{\bu\leftarrow \be}\dot\be\rangle}_{\text{grounded}} + \underbrace{\langle\partial_\bu\lum,\,\bJ_{\bu\leftarrow \bxi}\dot\bxi\rangle}_{\text{ungrounded}},
\end{equation}
where each term splits over actions as above, Fig.~\ref{fig:routes}. Alongside valence runs the direct, unsignalled term, $\langle\partial_\be\lum,\dot\be\rangle$, which moves the scale but is not felt. A stimulus that causes or reliably predicts a change in an essential variable signals it, and its drive is counted in the grounded term, as a noxious stimulus signals tissue loss.

A change \textbf{benefits} $\cS$ if, over the window, it moves $\cS$ away from $Z$ through its essential variables, relative to its absence, and \textbf{harms} it if toward it. How the change is \emph{felt} is set by the internal inputs. The two agree when essential variables affect internal inputs faithfully and internal inputs affect activity globally, as in the energy loop of \S\ref{ssec:dial}. They can come apart in two ways. An \textbf{ungrounded} change is felt though no essential variable has changed: phantom pain driven without damage, false alarms. Not every phantom is ungrounded. The cells that carry a signal are tissue too, so damage to them, as in neuropathic pain or a neuroma in an amputation stump, drives the signal: the damage is real and signalled, but felt where the signal reports, in the limb, rather than where it occurred. An \textbf{unsignalled} change moves the system toward $Z$ but is not felt: damage without nociception is not withdrawn from, and accumulates; a reserve whose signal is blocked runs down unnoticed. Because internal inputs can be driven directly, feelings can be induced without cause: opioids quiet the signal, so relief is felt while the damage continues.

\paragraph{Comments on the hypothesis.}
Interoception is necessary for valence, and also sufficient within a self-sustaining subject: any motion of the scale due to its internal inputs is felt, however faintly and whatever drives the inputs. The hypothesis fails if valence is felt in a system that is not a self-sustaining subject; if motion toward $Z$ that passes through no internal input is felt, for example if damage that is not signalled hurts; if an \emph{experience invert} exists, a system that feels negative valence as good, a valence analogue of the philosophical zombie~\citep{goff2025}; or if felt intensity does not rise with $|v|$. The hypothesis is supported abductively: the systems we think feel good or bad are self-sustaining, and affective circuitry is built on homeostatic regulation~\citep{denton2009,panksepp2005,damasio2013,solms2019}.

Valence is a rate, so a signal held constant is not felt: a sustained feeling is sustained motion of the felt terms, which unfelt terms of Eq.~\eqref{eq:vfull} (\S\ref{ssec:body}) can offset, so the scale itself need not fall. Weak tonic pain adapts, while stronger tonic pain intensifies~\citep{weissmanfogel2015}, as sensitisation keeps the gain of the pain pathway moving~\citep{woolf2011}. Pain that persists while its signals hold constant would count against the hypothesis.

\begin{table*}[t]
\centering
\small
\caption{\textbf{Phenomena the account explains.} Each phenomenon and the component of valence that accounts for it.}
\label{tab:phen}
\begin{tabular}{@{}>{\raggedright\arraybackslash}p{0.3\textwidth}>{\raggedright\arraybackslash}p{0.55\textwidth}>{\raggedright\arraybackslash}p{0.09\textwidth}@{}}
\toprule
Phenomenon & What accounts for it & Section \\
\midrule
& \textbf{The log scale} \\
The same food is pleasant when hungry, dull when sated (alliesthesia) & the same meal is a larger change of \emph{log} scale when the reserve is low & \S\ref{ssec:sign} \\
Hunger grows louder as reserves fall & a falling reserve moves the log scale faster near $Z$ & \S\ref{ssec:sign} \\
Metabolism is barely felt while reserves are ample & while a constraint is slack the gain barely moves & \S\ref{app:constraint} \\
Negativity bias & $Z$ is a floor at $-\infty$: the log scale is concave, so a loss moves it more than an equal gain & \S\ref{ssec:sign} \\
\midrule
& \textbf{The split over locations, Eq.~\eqref{eq:locate}} \\
Hunger is diffuse, pain localised & a global gain has no residual field; a local gain concentrates it & \S\ref{sec:hypV}, \S\ref{ssec:body}, \S\ref{ssec:pain} \\
Intensity grows with spread & each action counts in proportion to its share, $w_i$ & \S\ref{sec:hypV} \\
Hurt tracks what an injury costs & a learned global gain signals local damage to the whole body & \S\ref{ssec:learned} \\
\midrule
& \textbf{What passes through internal inputs, Definition~\ref{def:valence}} \\
Anaesthesia is not aversive & the drug lowers the scale directly, through no internal input & \S\ref{sec:hypV} \\
Congenital insensitivity to pain & damage moves the scale only directly, so it is not felt while its harm accumulates & \S\ref{ssec:pain} \\
Pain asymbolia & damage affects receptors, but their signal does not move the scale & \S\ref{ssec:pain} \\
Brain damage is mostly unfelt; central pain when it hits the pain pathway & lost brain tissue lowers the scale directly; a lesion of the pathway drives or cuts the signal & \S\ref{sec:hypV} \\
Depletion without hunger & the reserve signal is blocked & \S\ref{ssec:reading} \\
\midrule
& \textbf{The split over causes, Eq.~\eqref{eq:split}} \\
Dread before a threat, nocebo; relief when it fails to arrive & a forecast is an internal input: the scale moves with the forecast, and back when it is withdrawn & \S\ref{ssec:allo} \\
Phantom limb pain & in part, internal inputs driven with no change in the essential variable they signal & \S\ref{sec:hypV}, \S\ref{ssec:pain} \\
Analgesia without repair & the signal is quieted directly while the damage continues & \S\ref{sec:hypV} \\
Hunger with an undepleted reserve & the reserve signal is driven & \S\ref{ssec:reading} \\
Hunger suppresses pain due to receptors & hunger, through the reserve signal, eases the global gain; the cost of the injury is felt through the reserve instead & \S\ref{ssec:learned} \\
\botrule
\end{tabular}
\end{table*}

\section{Results}
\label{sec:results}

\begin{figure*}[t]
  \centering
  \resizebox{\textwidth}{!}{\begin{tikzpicture}[x=1cm,y=1cm, font=\small, line cap=round, line join=round,
  lab/.style={font=\scriptsize, inner sep=1pt, align=left},
  lead/.style={thin, metagray!80},
  ar/.style={-{Stealth[length=1.8mm]}, thick, metafg!70}]
\definecolor{essC}{HTML}{C08A1E}\definecolor{sigC}{HTML}{2C6FA8}\definecolor{falC}{HTML}{B03030}
\definecolor{skin}{HTML}{F3ECE6}\definecolor{edge}{HTML}{8A7F78}
\foreach \a/\b in {{(2.35,3.35)}/{(1.05,2.05)},{(3.65,3.35)}/{(5.05,2.3)},{(2.65,1.25)}/{(2.35,0.05)},{(3.35,1.25)}/{(3.65,0.05)}}{
  \draw[edge, line width=15pt] \a -- \b; \draw[skin, line width=13pt] \a -- \b;}
\draw[edge, line width=13pt] (3,3.5) -- (3,3.95); \draw[skin, line width=11pt] (3,3.5) -- (3,3.95);
\draw[edge, fill=skin, line width=1pt] (3,2.3) ellipse (0.95 and 1.35);
\draw[edge, fill=skin, line width=1pt] (3,4.35) circle (0.52);
\draw[metafg!40, fill=white] (3,4.4) ellipse (0.36 and 0.27);
\foreach \p in {(2.82,4.47),(3.0,4.33),(3.18,4.5),(2.92,4.6),(3.12,4.36)} \fill[metafg!60] \p circle (0.9pt);
\draw[metafg!40, thin] (2.82,4.47)--(3.0,4.33)--(3.18,4.5)--(2.92,4.6)--(2.82,4.47) (3.0,4.33)--(3.12,4.36);
\draw[essC, line width=0.9pt, rounded corners=3pt, fill=white] (2.6,1.55) rectangle (3.4,2.55);
\fill[essC!55, rounded corners=2pt] (2.65,1.6) rectangle (3.35,2.12);
\foreach \y in {1.72,1.95,2.18} \draw[essC!80, thick, -{Stealth[length=1.3mm]}] plot[domain=0:0.42, samples=20, smooth] ({3.47+\x},{\y+0.045*sin(\x*1700)});
\draw[sigC, very thick, densely dotted] (2.75,2.55) to[out=105,in=-115] (2.82,4.15);
\fill[essC!45] (4.4,2.8) circle (0.19);
\foreach \p in {(4.25,2.73),(4.5,2.9),(4.55,2.7)} \fill[sigC] \p circle (1.2pt);
\draw[sigC, very thick] (4.32,2.86) to[out=150,in=-30] (3.6,3.42) to[out=150,in=-55] (3.2,4.15);
\fill[falC] (3.85,3.3) circle (1.7pt);
\draw[metafg!80, fill=yellow!75!orange, line width=0.4pt] (5.25,3.25) -- (4.92,2.95) -- (5.05,2.93) -- (4.78,2.6) -- (5.2,2.98) -- (5.07,3.0) -- cycle;
\draw[essC!80!black, fill=essC!25] (0.6,1.45) circle (0.17); \draw[essC!80!black] (0.6,1.62) -- (0.66,1.75);
\draw[ar] (0.82,1.5) to[out=-5,in=200] (2.6,1.8);
\node[lab, anchor=east] (lnet) at (1.9,4.7) {network $f_\cW$: content $y_i$};
\draw[lead] (lnet.east) -- (2.66,4.45);
\node[lab, anchor=east, text=sigC] (lsig) at (1.9,3.55) {reserve signal $\hat E$\\forecast $\tau$, hunger $\eta$};
\draw[lead] (lsig.east) -- (2.74,3.3);
\node[lab, anchor=east, text=essC!70!black] (lres) at (1.75,2.55) {reserve $E$};
\draw[lead] (lres.east) -- (2.6,2.2);
\node[lab, anchor=north] at (0.6,1.22) {food};
\node[lab, anchor=north] at (1.75,1.42) {forage $\gamma$};
\node[lab, anchor=west, text=essC!70!black] (lleak) at (4.15,1.15) {leak $\kappa$, costs $C$\\[-1pt](burned as heat)};
\draw[lead] (lleak.west) -- (3.75,1.7);
\node[lab, anchor=west] (lhaz) at (5.45,3.25) {hazard $z_i$, load $L_i$};
\node[lab, anchor=west, text=essC!70!black] (ltis) at (5.45,2.62) {tissue $h_i$, repair $\varepsilon$};
\draw[lead] (ltis.west) -- (4.6,2.75);
\node[lab, anchor=west, text=sigC] (lrec) at (5.45,2.05) {receptors $r_i$; $\nu$, $\omega$};
\draw[lead] (lrec.west) -- (4.52,2.68);
\node[lab, anchor=west, text=falC] (lxi) at (4.15,3.95) {drive $\xi_i$};
\draw[lead] (lxi.west) -- (3.88,3.36);
\begin{scope}[shift={(8.0,0)}]
\node[anchor=west, font=\normalsize] at (0,4.35) {$\tilde y_i \;=\; {\color{sigC} m(\hat E)}\; {\color{metagray} q(E)}\; {\color{essC!80!black} h_i}\; {\color{sigC} G_i(\br)}\; y_i$};
\node[lab, anchor=west] at (0.05,3.75) {activity at location $i$: content scaled by four gains};
\node[lab, anchor=north west, text width=6.3cm] at (0,3.35) {\textcolor{sigC}{\rule{5pt}{5pt}}\; \textbf{felt}: gains set by internal inputs, the reserve signal and the receptors;\\ \phantom{\rule{5pt}{5pt}}\; $\log G_i = a\,r_i + (b+c\eta)\textstyle\sum_j r_j$};
\node[lab, anchor=north west, text width=6.3cm] at (0,2.35) {\textcolor{essC}{\rule{5pt}{5pt}}\; \textbf{not felt}: essential variables acting directly, the reserve through $q$ and tissue};
\node[lab, anchor=north west, text width=6.3cm] at (0,1.55) {\textcolor{falC}{\rule{5pt}{5pt}}\; \textbf{ungrounded}: a drive on the receptors with no damage behind it, felt through the gain};
\end{scope}
\end{tikzpicture}}
  \caption{\textbf{Model~\ref{m:full}: a body that sustains itself.}
    Biological counterparts: $E$ glucose, glycogen, fat, with leak $\kappa$ basal metabolism and costs $C$ exertion and repair; $\hat E$ glucose and hormonal sensing, or an allostatic forecast $\tau$ steps ahead; $m$ energy available for neural signalling; $q$ energy available locally; $h_i$ tissue integrity, $\varepsilon$ healing; $z_i$ a noxious stimulus, $L_i$ use of the part; $r_i$ nociceptors, $\nu$ inflammatory drive, $\xi_i$ a drive without damage, e.g. cortical reorganisation after amputation, $\omega$ the withdrawal reflex; $a$ guarding, $b$ whole-body rest; $\eta$ hypothalamic hunger signals, with weight $c$.}
  \label{fig:body}
  \figalttext[Schematic of the model body, with one injured location drawn.]{The energy reserve, its signal or forecast, the metabolic gain, local energy, tissue integrity, noxious input, receptors, withdrawal and guarding, and hunger signals, with arrows showing how each affects the others and the activity at one location, an injured arm.}%
\end{figure*}

Hypothesis~\ref{h:valence} is applied to a differentiable model of a body that sustains itself, \S\ref{ssec:body}. The motion of its scale splits into six terms, Eq.~\eqref{eq:vfull}: the reserve signal, the reserve's direct effect, tissue, the local and global parts of the receptor gain, and content. Three are due to internal inputs, and are valence.

Experiments are constructed to illustrate terms of Eq.~\eqref{eq:locate} or Eq.~\eqref{eq:split}; their role is to show that the decompositions yield the phenomena of Table~\ref{tab:phen}. Experiments marked \emph{(check)} sanity-check an expected consequence. Experiments~\ref{exp:alli}, \ref{exp:reading} and~\ref{exp:pain} follow from the definitions and choices made by hand. In Experiment~\ref{exp:allo} the forecast is fitted, and in Experiment~\ref{exp:allo_online} it is learned online. In Experiments~\ref{exp:learned} and~\ref{exp:hunger} the form of the gain is given and its sign and size are learned, as is the policy in Experiment~\ref{exp:constraint}.

If activity is a gain times a network's output, $\tilde y=g\,f_\cW(\bx)$, then $\lum=\log g+\log\|\nabla_\cW f_\cW\|$: the gain says how far the system can act, and the network, its \emph{content}, what it responds to and how. Both terms can move the scale. In general essential variables can affect content $f_\cW$, which is then an internal input. However, in Model~\ref{m:full} for simplicity the net only responds to external inputs $\bx$, so its motion is not felt, and only the gains matter for valence. Each gain stands for a physiological quantity, such as the energy available for signalling, intact tissue or a receptor's drive. Content gives the Jacobian its shape, what the experience is about (Hypothesis~\ref{h:first}).

Except in \S\ref{ssec:learned}, the network $f_\cW$ of Model~\ref{m:full} is a recurrent network with scalar input and 16 hidden units. Where parameters are learned (\S\S\ref{ssec:allo}, \ref{ssec:learned}, \ref{app:constraint}), valence is reported within episodes, at fixed parameters. Results are over ten random systems. The model is far from biology, so its results matter \emph{qualitatively}.

\subsection{A model of a self-sustaining body}
\label{ssec:body}

An organism sustains many essential variables, which affect its activity through internal inputs. Model~\ref{m:full} has two kinds of essential variable, an energy reserve shared by the whole body and tissue at each location, and four gains on activity: two set by internal inputs, the reserve signal and the receptors, and two direct, the reserve's own effect on activity and tissue itself. Fig.~\ref{fig:body} shows its structure and notation.

\begin{model}
  \label{m:full}
  A body has $n$ locations. A neural network $f_\cW$ produces activity $y_{i,t}=f_\cW(\bx_{\le t})_i$ at each location: what the body would do in optimal conditions. The body's actual activity is
  \begin{equation}
    \label{eq:full}
    \tilde y_{i,t} = m(\hat E_t)\; q(E_t)\; h_{i,t}\; G_i(\br_t)\; y_{i,t}.
  \end{equation}
  \textbf{Essential variables.} A shared leaky reserve $E_t$ is replenished by activity, and pays costs,
  \begin{equation}
    \label{eq:energy}
    E_t = (1-\kappa)E_{t-1} + \gamma\, E_{\rm repl}(\tilde\by_{t-1}) - C_t,
  \end{equation}
  with leak $\kappa$, yield $\gamma$ and a replenishment $E_{\rm repl}$ that increases with activity, $E_{\rm repl}(\bm 0)=0$; and tissue $h_{i,t}\in[0,1]$ at each location, which repairs slowly and is lost under a hazard $z_{i,t}$ in proportion to the load $L_{i,t}$ that activity places on it,
  \begin{equation*}
    h_{i,t} = h_{i,t-1} + \varepsilon(1-h_{i,t-1}) - z_{i,t}\,L_{i,t}.
  \end{equation*}
  Low energy also slows activity directly, by $q(E)=E/(E+E_q)$.\\
  \textbf{Internal inputs.} A metabolic gain $m(\hat E)=\hat E/(\hat E+E_0)$, set by the system's signal $\hat E_t$ of its reserve, a forecast $\tau$ steps ahead ($\tau=0$: $\hat E_t=E_t$); receptors, silenced by withdrawal $\omega_t\in[0,1]$,
  \begin{equation*}
    r_{i,t} = (1-\omega_t)\big(z_{i,t}+\nu(1-h_{i,t})+\xi_{i,t}\big);
  \end{equation*}
  and a receptor gain with a local and a global part, the global part open to the hunger signal,
  \begin{equation*}
    \log G_i(\br) = a\,r_i + (b+c\,\eta)\textstyle\sum_j r_j,
  \end{equation*}
  with the hunger signal $\eta=m(E_{\rm sat})-m(\hat E)$, a second signal of the reserve.
\end{model}

\noindent
The gains are $m$, $q$, $h$ and $G$. The direct factor $q$ is switched off ($E_q=0$, so $q\equiv1$) except in \S\ref{ssec:reading}; in a body $q$ acts in each local store, while all of them affect $\hat E$. $C_t$ is the cost of an action, or of repairing tissue. A negative cost is a meal. The term $\xi_{i,t}$ is a drive on the receptors with no damage behind it, zero except in the phantom conditions. The hunger signal $\eta$ is zero when the reserve is at its sated level $E_{\rm sat}$ and positive below it; its weight $c$ is zero except in Experiment~\ref{exp:hunger}. There the body is called hungrier when scarcity lowers its reserve and raises $\eta$; the feeling of hunger is the valence produced as the reserve signal moves.

The model has two loops, one for each essential variable. In the \emph{energy loop}, activity replenishes the reserve, whose signal sets the metabolic gain; in the \emph{tissue loop}, damage drives the receptors, which protect tissue through withdrawal and guarding. The model is self-sustaining when either loop is closed: with the actions cut the reserve leaks, and under hazard tissue is at risk. The experiments close the energy loop (\S\S\ref{ssec:dial}--\ref{ssec:allo}), then the tissue loop (\S\ref{ssec:pain}), then both (\S\ref{ssec:learned}). The forecast is learned in \S\ref{ssec:allo}. The receptor gain is set by hand until \S\ref{ssec:learned}, where it is learned. In every case the body is one subject by construction, since every row of $\bJ_\cS$ depends on the same content parameters.

\paragraph{Valence in Model~\ref{m:full}.}
Here $\bJ_\cS$ is evaluated with the state variables held at their current values: it is the Jacobian of each step's actions, not differentiated through the history of the reserve or tissue. Row $(t,i)$ of $\bJ_\cS$ is $m(\hat E_t)\,q(E_t)\,h_{i,t}\,G_i(\br_t)$ times a row set by content $f_\cW$, of norm $k_{i,t}$, so $\tfrac{d}{dt}\log\ell_i=\tfrac{d}{dt}\log m+\tfrac{d}{dt}\log q+\tfrac{d}{dt}\log h_i+\tfrac{d}{dt}\log G_i+\tfrac{d}{dt}\log k_i$. Since the shares sum to one, the split over actions of Eq.~\eqref{eq:locate}, applied to $\dot\lum$, gives
\begin{equation}
  \label{eq:vfull}
  \begin{split}
  \dot\lum={}&\underbrace{\tfrac{d}{dt}\log m(\hat E)}_{\text{reserve signal}}
  +\underbrace{a\textstyle\sum_i w_i\,\dot r_i}_{\text{local gain}}
  +\underbrace{\tfrac{d}{dt}\big[(b+c\eta)\textstyle\sum_j r_j\big]}_{\text{global gain}}\\
  &+\underbrace{\tfrac{d}{dt}\log q(E)}_{\text{reserve, direct}}
  +\underbrace{\textstyle\sum_i w_i\,\tfrac{d}{dt}\log h_i}_{\text{tissue}}
  +\underbrace{\textstyle\sum_i w_i\,\tfrac{d}{dt}\log k_i}_{\text{content}}.
  \end{split}
\end{equation}
The first three terms are due to internal inputs, the reserve signal and the receptors: they are valence, $v$. The rest are not felt: essential variables acting directly, and content, which moves only with external inputs. The first, third and fourth terms move every action alike, so they add nothing to the residual field: they are diffuse. The local gain and tissue terms are localised, each location's change diluted by its share; of these only the local gain is felt, so it alone localises valence. 

With $c=0$, as everywhere except Experiment~\ref{exp:hunger}, the global term is $b\sum_j\dot r_j$, and hunger lives in the first term alone. In discrete time each derivative is a difference across a step, with midpoint shares, and cumulative valence is $\sum_t v_t$.

Each felt term splits by cause, Eq.~\eqref{eq:split}. In the reserve term, the reserve acts through its signal $\hat E$, or a forecast that a cue can falsely drive. In the receptor terms, tissue acts ($z$, $h$), or the drive is ungrounded ($\xi$).

\subsection{Open and closed loops}
\label{ssec:dial}

Experience requires useful energy~\citep{attwell2001,montague2006,niven2016,quintela2022}; as the supply falls to zero, activity and responsiveness fall with it.

Switch off everything in Model~\ref{m:full} but the metabolic gain, on a single action: no tissue ($h\equiv1$), no receptors ($G\equiv1$), no direct factor ($q\equiv1$), and $m_t\in[0,1]$, how much useful energy the system has available, set from outside,
\begin{equation}
  \label{eq:metabolism}
  \tilde y_t = m_t\, y_t .
\end{equation}
Moving $m$ is radial motion in Jacobian space (Fig.~\ref{fig:affect}): toward or away from $Z$, changing scale without changing shape.

Here, $m$ is an externally controlled dial, so it is not an internal input, and by Hypothesis~\ref{h:valence} it has no valence: its experience may be amplified or diminished, but there is no experience of amplification or diminishment. An example is a patient under anaesthesia (\S\ref{sec:hypV}).

The closest biological counterpart to $m$ is the energy supplied for neural signalling, which is most of the brain's energy budget~\citep{attwell2001,harris2012}; when supply fails, responsiveness falls and consciousness is lost.

\paragraph{Closing the loop.}
Switch on the reserve, Eq.~\eqref{eq:energy}, with no costs and $\tau=0$: the gain is $m_t=m(\hat E_t)$ with $\hat E_t=E_t$, and the system's actions now change it. This closes the \emph{energy loop}. The system's actions affect $m$, and $m$ scales its Jacobian (Fig.~\ref{fig:body}). The reserve is an essential variable, and $\hat E$ the system's signal of it, an internal input through which it affects activity everywhere, via $m$; the motion of $m$ is felt. The signal is a variable of its own that here happens to equal the reserve; \S\ref{ssec:reading} pulls the two apart. The energy loop is self-sustaining: without replenishment the reserve decays at rate $\kappa$ and the system drifts toward $Z$; its actions replenish the reserve, and $\partial_{\hat E}\lum = m'(\hat E)/m(\hat E) > 0$, so they move it away. At an equilibrium of the closed loop, the reserve holds steady and $v=0$.

The energy loop, as modelled, is a recurrent network simulated on a computer. The computer is not self-sustaining because it is externally powered: the hypothesis applies to the system being modelled, not to the simulation.

\subsection{Depletion and need}
\label{ssec:sign}

In the energy loop only the first term of Eq.~\eqref{eq:vfull} is on, so $v=\tfrac{d}{dt}\log m$, and depletion raises intensity for two reasons. The first is the logarithm (\S\ref{sec:hypV}): the same meal is a larger proportional change to a small reserve, so it moves $\lum$ further. The second is that $m$ saturates: when nearly full, more energy changes little, and the reserve's own decay moves $\log m(E)$ at rate $\kappa E_0/(E+E_0)$, faster as $E$ falls, though never faster than the leak itself, since $E_0/(E+E_0)<1$. So \textbf{hunger grows louder as reserves fall, and a meal is most welcome when most needed}.

\experiment{exp:alli}{Alliesthesia (check)} The same meal, delivered at different reserve levels, always moves the system away from $Z$, and its intensity, its effect on $\lum$ relative to the same run without it, falls by more than an order of magnitude from a nearly empty reserve to a full one. It tracks the change in $\log m$; here the Jacobian is taken through the whole loop.

\paragraph{Positive affect and negativity bias.}
Three features of valence in Model~\ref{m:full} make negative affect more prominent: $Z$ is a floor with no matching ceiling; on the log scale a loss moves the system further than an equal gain; and the saturating $m$, a modelling choice, makes gains silent when replete. Positive affect is then relief scaled by need, as in homeostatic reinforcement learning~\citep{keramati2014}. \textbf{Negative affect is anchored to a hard floor, and so is sharp and urgent; positive affect, as relief, is bounded and scaled by need}, which matches negativity bias~\citep{baumeister2001,rozin2001}.

\paragraph{Metabolism as a constraint.}
Most of the time metabolism is barely felt: dedicated circuitry implementing drives keeps humans clear of the regime where energy limits neuronal responsiveness~\citep{andermann2017}. When a learned policy spends the reserve subject to a metabolic constraint, valence is close to zero while the constraint is slack, and several times larger, and negative, once it binds (Experiment~\ref{exp:constraint}, \S\ref{app:constraint}). \textbf{Metabolic valence is quiet while the system is comfortable, and loud once it has spent down to its constraint.}

\subsection{Blocking and driving the reserve signal}
\label{ssec:reading}

So far the signal is the reserve, so the two cannot disagree. In a body they can: energy is held in many local stores, and a quantity that depends on stores everywhere and acts on activity everywhere has to act through a signal of its own, which can fail or be driven. Hypoglycaemia unawareness, in which people with long-standing diabetes lose the warning symptoms of low blood glucose, and the hunger induced by hormones that signal an empty stomach, are concrete examples~\citep{cryer2004,wren2001}.

Take the energy loop and switch on the direct factor $q$, so that low energy also slows activity where it occurs, and raise the yield so that the closed loop still holds a reserve away from $Z$. The reserve now moves the scale by two routes: through its signal, $m(\hat E)$, which is felt, and directly, through $q(E)$, which is not.

\begin{figure}[t]
  \centering
  \includegraphics[width=\columnwidth]{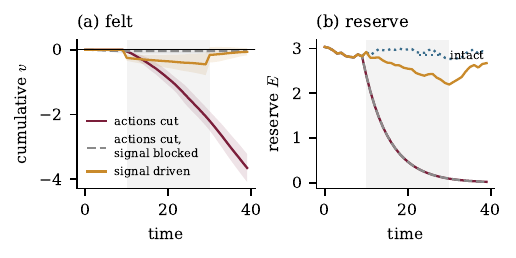}
  \caption{\textbf{Hunger needs a signal.} Experiment~\ref{exp:reading}; actions cut from the start of the shaded window, or the signal driven during it.
  \textbf{(a)} Cumulative valence, due to the reserve signal, relative to the same run left intact (blocked: absolute, since the signal does not move).
  \textbf{(b)} The reserve. With the signal blocked (dashed), the reserve falls exactly as when unblocked, but nothing is felt.}
  \label{fig:reading}
  \figalttext[Line plots of cumulative valence and of the reserve when the reserve signal is cut, blocked or driven.]{Left: cumulative valence due to the reserve signal over time, falling when actions are cut, staying at zero when the signal is blocked, and dipping then recovering when the signal is driven. Right: the reserve falls the same way whether the signal is blocked or not.}%
\end{figure}

\experiment{exp:reading}{Hunger needs a signal} Start each system at the closed loop's steady state and, at one step, cut its actions off from the reserve. The reserve runs down toward $Z$, and the fall is felt through its signal, increasingly as the reserve empties (Fig.~\ref{fig:reading}a). Block the signal as the actions are cut, holding $\hat E$ at its last value: the reserve runs down exactly as before (Fig.~\ref{fig:reading}b), and the scale falls with it through $q$, but valence is zero throughout. \textbf{Depletion without a signal is not felt.} Conversely, drive the signal to half the reserve, with the actions intact: valence falls at once, and recovers when the drive stops, though the reserve starts at its steady state. The lowered gain also cuts foraging, so the reserve does fall: ungrounded hunger has real consequences through the loop. \textbf{A driven signal is felt as hunger with an undepleted reserve.}

\subsection{Adding a forecast}
\label{ssec:allo}

Affect often anticipates changes in essential variables~\citep{sterling2012,kleckner2017}: hunger arrives at the usual mealtime, and dread before the threat. To model allostasis, take the energy loop and switch on the forecast of Model~\ref{m:full}: the metabolic gain is set by a learned forecast $\hat E_t$ of the reserve $\tau$ steps ahead, rather than by $E_t$. The forecast is an internal input: the loop acquires a route from actions, through $\hat E$, onto current scale, and the reserve term of Eq.~\eqref{eq:vfull} moves before the reserve does.

Allostatic valence is the valence of \emph{predicted} motion relative to $Z$; homeostatic valence is the special case $\tau=0$ where the current state is the forecast. When forecast and reserve disagree, valence should follow the forecast first and the reserve once the error is absorbed. When a forecast is revised by its errors, valence is close to a prediction error, as in accounts on which valence tracks prediction error or the rate of change of free energy~\citep{joffily2013,hesp2021}. The accounts differ in what moves: here only predicted motion of the scale relative to $Z$ counts, so the same forecast error is felt more strongly the closer the system is to $Z$.

Dread and hunger both lower the scale, but differ in their causes: hunger is due to changes in the reserve and dread due to forecast changes. Valence due to a forecast should yield to information that corrects it, such as reassurance, with no change in the reserve, while hunger from depletion yields only until its signal is corrected against the reserve (Table~\ref{tab:pred}, row~3).

\begin{figure*}[t]
  \centering
  \includegraphics[width=\textwidth]{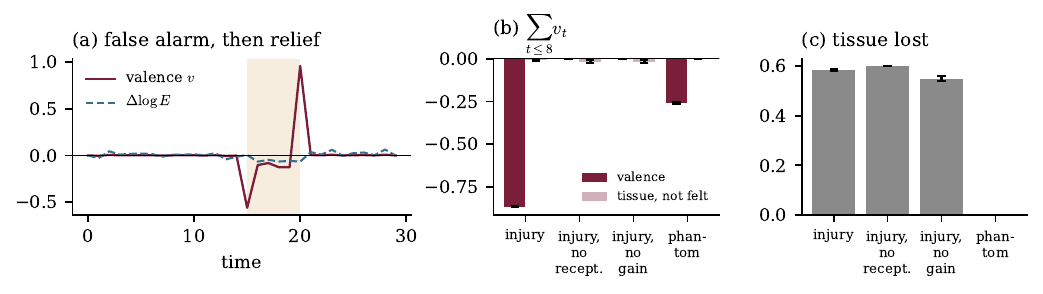}
  \caption{\textbf{Allostatic valence leads the reserve; pain dissociates as in the clinic.}
    \textbf{(a)} Experiment~\ref{exp:allo} with a false cue (shaded): valence $v$, dominated by the change in gain set by the forecast, falls at the cue and turns positive when the cue ends, while the reserve barely moves (dashed: its per-step change, $\Delta\log E$).
    \textbf{(b, c)} Experiment~\ref{exp:pain}, 32 locations: an injury to the intact body, with receptors off, or with the gain off, and receptors driven without injury (phantom): 
    \textbf{(b)} cumulative valence from the start of the episode to the end of the damage, $\sum_t v_t$ (dark), and the motion of the scale due directly to tissue over the same steps, which is not felt (light);
    \textbf{(c)} tissue lost at the damaged location, at the end of the episode.}
  \label{fig:capres}
  \figalttext[Plots of anticipatory valence with a false cue, and bar charts of pain dissociations.]{Left: valence falls at a false cue and turns positive when the cue ends while the reserve barely moves. Middle: cumulative valence over an injury for the intact body, without receptors, without the gain, and with receptors driven without damage, beside the unfelt direct effect of tissue loss. Right: tissue lost at the damaged location in each condition.}%
\end{figure*}

\experiment{exp:allo}{Anticipation and relief (check)} Suppose a cue precedes a drain on the reserve by five steps, and the forecast, which looks five steps ahead, is fitted to predict the drain from the cue. With the forecast on, valence turns negative at the cue in all ten systems, and so five steps before the homeostatic case ($\tau=0$) responds to the drain itself. When the cue is false, valence falls at the cue and turns positive when it ends, though the reserve barely moves: a false alarm, and then relief when the expected threat fails to arrive, Fig.~\ref{fig:capres}a.

\begin{figure}[t]
  \centering
  \includegraphics[width=\columnwidth]{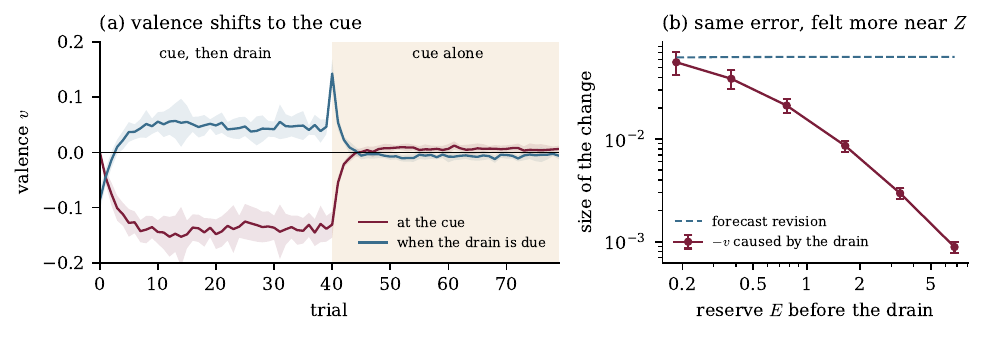}
  \caption{\textbf{Valence shifts to the cue that predicts the drain; the same error is felt more near $Z$.} 
  Experiment~\ref{exp:allo_online}. 
  \textbf{(a)} Valence at the cue and when the drain is due, as the cue is paired with the drain, then without it (shaded). 
  \textbf{(b)} The same unexpected drain at different reserve levels: the forecast error is always the same, but it is felt more strongly at lower reserves.}
  \label{fig:online}
  \figalttext[Plots showing dread moving to a predictive cue over trials, and the same error felt more strongly at low reserve.]{Left: over trials of a cue followed by a drain, valence moves from the drain to the cue, and fades again when the cue stops predicting the drain. Right: the same unexpected drain is felt more strongly the lower the reserve.}%
\end{figure}

\experiment{exp:allo_online}{Learning the forecast} Let the forecast learn the cue online, from nothing, by a delta rule on the weight of the cue. Over trials, valence shifts to the cue that predicts the drain, Fig.~\ref{fig:online}a, much as dopamine responses move from a reward to the cue that predicts it~\citep{schultz1997}; when the cue stops predicting a drain, the dread fades. The model predicts how the same error is felt in different conditions: a forecast revision of fixed size is felt much more strongly at a low reserve than at a high one, Fig.~\ref{fig:online}b. \textbf{Valence follows the forecast's motion relative to $Z$, not its error.}

\subsection{Adding tissue and receptors: pain}
\label{ssec:pain}

Pain is localised, urgent and hard to ignore. It can mislead: there is pain without damage (phantom limb pain) and damage without pain (congenital insensitivity).

Switch the reserve off ($m\equiv1$) and switch tissue, receptors and the receptor gain on, with $n$ locations and the gain set by hand: $a=-\alpha$ and $b=-\beta$, so that
\begin{equation*}
  G_i(\br) = \exp\Big(-\alpha\, r_i - \beta \textstyle\sum_{j} r_j\Big)
\end{equation*}
suppresses activity everywhere, and most at the damaged location. A reflex withdrawal $\omega_t$, driven by the receptors, removes the load, $L_{i,t}=1-\omega_{t-1}$. This closes the \emph{tissue loop}. Tissue requires effort to sustain: damage accumulates unless repaired, a damaged part contributes less to $\bJ_\cS$ because it is less responsive, and withdrawal and guarding affect how damage progresses; the tissue loop is self-sustaining. The gain here models the recuperative phase of pain, guarding and rest, in which responsiveness narrows and falls~\citep{wall1979,bolles1980}. Much of the arousal around an acute injury belongs instead to the defensive response, which, like hunger in Experiment~\ref{exp:hunger}, suppresses pain~\citep{bolles1980}. The scale measures the breadth of responsiveness, not arousal.

Pain is both localised and global. It is localised because tissue affects receptors location by location and the local part of the gain changes most at the damaged one, concentrating the residual field there; hunger, in the reserve term, has no residual field. It is global because the global part of the gain changes activity everywhere, so the whole scale moves. \textbf{Receptors sense damage and broadcast it}: without them, only the tissue term is left, which passes through no internal input, and the damage is not felt.

\experiment{exp:pain}{Dissociations of pain} 
The tissue loop with 32 locations, damage at one, and reflex withdrawal. The gain suppresses the damaged location, and other locations to a lesser extent. In each condition, compare the cumulative valence from the start of the episode to the end of the damage, $\sum_t v_t$, with the damage actually taken. In the intact body, cumulative valence is negative in every system, and the residual field $\delta$ is negative and concentrated at the damaged location, as far as it can be, since $\delta$ sums to zero, and small at every other location. The damage is felt, especially where it occurs.

The valence comes entirely through the receptor gain, from the local and global gain terms of Eq.~\eqref{eq:vfull}. Damaged tissue dimming its own activity, the tissue term, also moves the scale toward $Z$, though by much less. It is not felt because it does not pass through an internal input (Fig.~\ref{fig:capres}b, light bars). Without receptors, valence is zero when tissue is damaged. More tissue is lost, because withdrawal no longer tracks the damage: damage without affect. Without the gain, receptors still fire at the damaged location and drive withdrawal, but their signal no longer affects the scale: nociception without affect. With receptors driven by $\xi$ and no damage, the drive is felt in every system, though no tissue is lost: affect without damage where it is felt, an ungrounded signal, Fig.~\ref{fig:capres}b,c.

These patterns correspond to congenital insensitivity to pain, pain asymbolia and phantom limb pain~\citep{nagasako2003,berthier1988,flor2006}. The first correspondence is exact: without receptors, damage is not felt at all, as in the clinic. The correspondence to asymbolia is the loosest: in the model withdrawal is driven by receptors and survives without the gain, whereas asymbolics often fail to withdraw~\citep{grahek2007}.

\subsection{Learning and pain}
\label{ssec:learned}

\begin{figure*}[t]
  \centering
  \includegraphics[width=\textwidth]{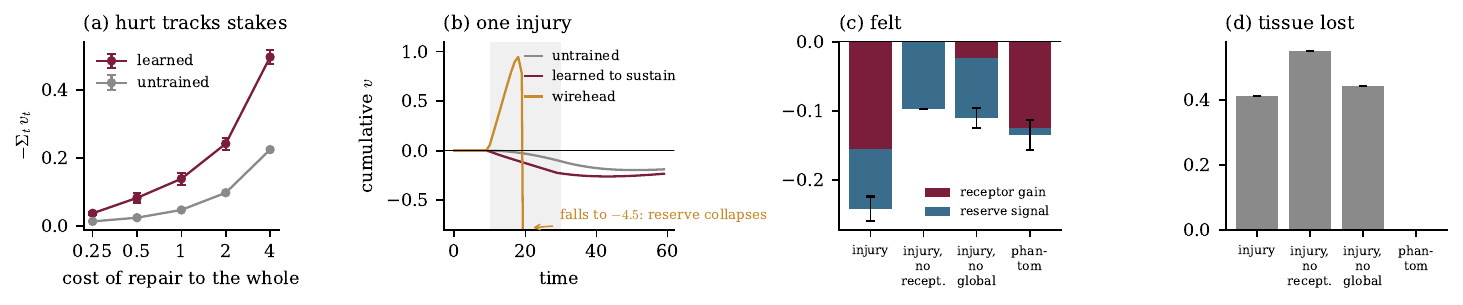}
  \caption{\textbf{A learned gain makes injuries hurt in proportion to cost.} Experiment~\ref{exp:learned}.
    \textbf{(a)} Intensity of one injury, $-\sum_t v_t$ over the injury, against the energy cost of repair, which sets how much a local injury matters to the whole: with the gain learned to sustain the body, and untrained ($a=b=0$).
    \textbf{(b)} Cumulative valence through one injury (shaded), one system: untrained; learned to sustain the body; and learned to increase the log scale (wirehead), which first rises and then falls far below the others as the reserve collapses.
    \textbf{(c, d)} The learned system: an injury to the intact body, with receptors off, or with the global part of the gain off, and receptors driven without injury (phantom): \textbf{(c)} cumulative valence over the injury, split by the internal input it passes through, the receptor gain or the reserve signal (error bars: s.d.\ of the total); \textbf{(d)} tissue lost at the injured location.}
  \label{fig:learned}
  \figalttext[Plots showing that a learned receptor gain makes injuries hurt in proportion to their cost.]{Panel a: injury intensity rises with the energy cost of repair when the gain is learned to sustain the body, but not when untrained. Panel b: cumulative valence through one injury for an untrained, a learned and a wirehead system, where the wirehead first rises and then falls far below. Panels c and d: valence split by internal input, and tissue lost, for the learned system under lesions and phantom drive.}%
\end{figure*}

In the tissue loop the gain was set by hand, choosing that nociception suppresses activity everywhere, and so how the damage is felt. Set this way it even costs the body: suppressing activity weakens withdrawal, so more tissue is lost. Would a system discover a better gain for itself?

Switch the shared reserve back on, alongside tissue and receptors, and let the gain be learned: this closes \emph{both loops} in the sense of Definition~\ref{def:loop}, coupled through the shared reserve, with every term of Eq.~\eqref{eq:vfull} on except the direct factor $q$. Activity forages, refilling the reserve, and loads any injury, both at the injured location and, through movement of the whole body, everywhere; repairing lost tissue draws on the reserve (details in \S\ref{app:learned}). The local part $a$ and global part $b$ of the gain are learned by gradient ascent on what benefits the body (\S\ref{sec:hypV}),
\begin{equation*}
  \textstyle\sum_t\log\big[m(E_t)\,\|\bh_t\odot\bk\|\big],
\end{equation*}
the log scale with the gain removed, where $\bk$ are the norms of the content's rows. The cost of repair sets how much an injury at one location matters to the body as a whole.

\experiment{exp:learned}{Learning to feel pain}
At the default cost of repair, the body learns to guard the injured part ($a<0$) and to rest as a whole ($b<0$). The more an injury costs the body, the stronger the global rest, and the more the injury hurts (Fig.~\ref{fig:learned}a). Because the global part lowers activity everywhere, the injury is now felt more strongly than untrained, and it is felt at once rather than as damage builds up (Fig.~\ref{fig:learned}b). The body also loses less tissue than when untrained. In short, \textbf{the learned gain makes an injury hurt in proportion to how much it matters}. That damage affects the whole body (\S\ref{ssec:pain}) is learned rather than assumed.

The same patterns reappear with both loops (Fig.~\ref{fig:learned}c,d), with one difference. Without receptors, the injury is still felt, though more weakly, and more tissue is lost. It is felt only through the reserve signal (Fig.~\ref{fig:learned}c): repair draws on the shared reserve, so the injury's cost is signalled as a fall in the reserve. In this body the reserve signal is itself a route that signals local damage to the whole; it is felt as hunger, not as pain. Without the global part of the gain, the injured part is still guarded, but the injury hurts about as much as unsignalled damage. Driving receptors with no damage is felt, less strongly than the injury, though no tissue is lost.

The objective leaves out the gain. Suppose the gain is instead learned to maximise the log scale itself, with the gain terms included: it learns to raise activity everywhere when injured. The injury then feels good at first, but soon the reserve collapses and the tissue is lost (Fig.~\ref{fig:learned}b). This is Goodhart's law~\citep{goodhart1975}: the learner optimises a signal, the gain, in place of the essential variables that affect it. The system is a wirehead that raises activity far beyond what the body can sustain. Learning is itself felt: \emph{learning to sustain the body feels bad, because it sensitises the body to injury}; learning toward the wirehead's objective feels strongly good.

The hurt comes from the response the gain drives, not from damage as such: the body learns to rest, and internally enforcing rest lowers the scale. When the best response to a load is more activity, and the body can sustain it, the load should be felt as less bad, or as good: exercise strains tissue yet is often experienced as good.

\experiment{exp:hunger}{Hunger suppresses pain due to receptors} Suppose the body of Experiment~\ref{exp:learned} faces scarcity that varies across episodes, lowering the reserve's equilibrium $E^*$ from its sated level, $E_{\rm sat}=1$, to $0.5$. Let the hunger signal $\eta$ affect the global part of the gain, with its weight $c$ learned. The term $c\,\eta$, by which the hunger signal reaches the global gain, is built into the model. It roughly corresponds to the circuit by which hunger-sensing hypothalamic neurons suppress inflammatory pain in mice~\citep{alhadeff2018}.

\begin{figure*}[t]
  \centering
  \includegraphics[width=\textwidth]{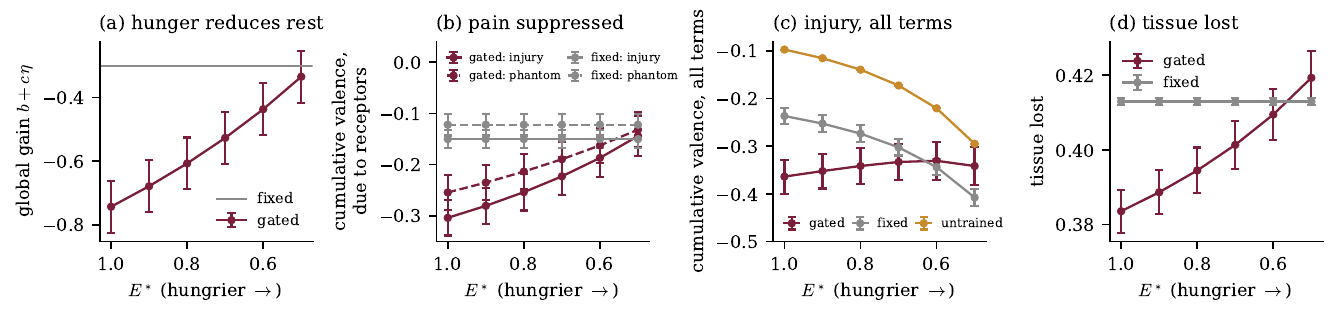}
  \caption{\textbf{Hunger suppresses pain due to receptors.} Experiment~\ref{exp:hunger}, against the reserve's equilibrium $E^*$ (hungrier to the right). Gated: the hunger signal affects the global gain, $c$ learned; fixed: $a$ and $b$ learned alone.
    \textbf{(a)} The learned global gain, $b+c\eta$, at the onset of injury.
    \textbf{(b)} Cumulative valence over the injury due to receptors: the gain terms without the hunger signal's part, $c\,\dot\eta\sum_jr_j$; dashed, the same for receptors driven without damage (phantom).
    \textbf{(c)} Cumulative valence over the injury, all felt terms, relative to the same run without injury; untrained: $a=b=c=0$.
    \textbf{(d)} Tissue lost at the injured location.}
  \label{fig:hunger}
  \figalttext[Plots showing that hunger suppresses pain due to receptors.]{Against the reserve's equilibrium, from sated to hungry: the learned global gain at injury onset, cumulative valence due to receptors falling with hunger, total felt valence over the injury, and tissue lost, for a gain that hunger can affect and one it cannot.}%
\end{figure*}

In all ten systems the learner sets $c>0$: the hungrier the body, the less it rests when injured (Fig.~\ref{fig:hunger}a). The global gain term now also includes the hunger signal's own motion, $c\,\dot\eta\sum_jr_j$, which is due to the reserve signal (splitting a product between its two inputs is a convention of the model); counting it there, the hurt due to receptors falls with hunger, and phantom pain falls too (Fig.~\ref{fig:hunger}b). Hunger does not make the injury feel better overall: the total barely changes, because the hurt moves from the receptors to the signals of the reserve, and the hungry body pays in tissue; with a gain that hunger cannot affect, the total instead grows with hunger (Fig.~\ref{fig:hunger}c,d). \textbf{Hunger suppresses the pain due to receptors, but not the hurt of the injury: its cost is felt instead through the reserve, as deeper hunger.} 

\begin{table*}[t]
\centering
\small
\caption{\textbf{Consequences for organisms.} Phenomena predicted by the account and where they are derived.}
\label{tab:pred}
\begin{tabular}{@{}p{0.02\textwidth}p{0.77\textwidth}>{\raggedright\arraybackslash}p{0.14\textwidth}@{}}
\toprule
& Consequence & Reference \\
\midrule
1 & \emph{Blocking a signal removes the affect, not the need; driving it creates the affect without the need.} Without nociception, damage is felt only through other signals, such as the energy repair draws (congenital insensitivity); without the gain, nociception has no affect (asymbolia); driven receptors give pain without damage (phantom limb pain). Without glucose sensing, depletion goes unfelt (hypoglycaemia unawareness); hormones signalling an empty stomach give hunger with full reserves. & Exps.~\ref{exp:reading}, \ref{exp:pain}, \ref{exp:learned} \\
2 & \emph{Valence follows the forecast, then the reserve.} Valence leads the reserve when the two differ, as in anticipatory nausea, nocebo and relief, and the same forecast error is felt more when depleted, as hunger amplifies dopamine responses to food cues~\citep{cone2016}. & Exps.~\ref{exp:allo}, \ref{exp:allo_online} \\
3 & \emph{Information moves valence through signals, not the reserve.} Reassurance dissolves dread with no change in the reserve; believing a meal larger than it is lowers ghrelin as a larger meal would~\citep{crum2011}. Hunger from depletion yields to information only until its signal is corrected against the reserve. & \S\ref{ssec:allo} \\
4 & \emph{Deafferentation blunts affect only where it cuts signals of essential variables.} Interrupting vagal or humoral signalling should blunt emotion; spinal cord injury, which spares them, does not~\citep{bermond1991,deady2010}; but see~\citet{hohmann1966}. & Exps.~\ref{exp:reading} (blocked), \ref{exp:learned} (receptors off) \\
5 & \emph{Hurt tracks the value of resting.} When rest is the best response, an injury hurts more, and more globally; when a competing need calls for activity, pain due to receptors is suppressed, as by hunger~\citep{alhadeff2018}, and the injury's cost is felt as deeper hunger. & Exps.~\ref{exp:learned}, \ref{exp:hunger} \\
6 & \emph{Intensity grows with spread.} A change in a signal is felt more the more of the system it moves, as larger stimuli hurt more~\citep{price1989}; a signal acting on one part is diluted by that part's share unless it is carried further. & \S\ref{sec:hypV}; Exp.~\ref{exp:learned} \\
7 & \emph{Lesion location, not size, predicts affect.} Damage to the pathways carrying internal inputs, such as thalamus, insula or cingulate, changes affect, as central pain or asymbolia; equal or larger damage elsewhere does not. & \S\ref{sec:hypV}; Exps.~\ref{exp:reading}, \ref{exp:pain} \\
8 & \emph{Valence is a rate.} The end of a pain or threat is felt as relief while the system is still worse off than before, and relief grows with what ends~\citep{leknes2008}. & Exps.~\ref{exp:reading} (driven), \ref{exp:allo} \\
\botrule
\end{tabular}
\end{table*}

\section{Discussion and conclusions}
\label{sec:discussion}

The paper makes three claims. First, valence is the velocity of the log scale due to the system's internal inputs: only the interoceptively registered motion of a system relative to its terminal set is felt, and valence is the experience of a rescaling of experience. Structural theories of consciousness, such as integrated information theory~\citep{tononi2016}, say what an experience is like but assign it no valence; here, valence is the motion of the same structure that shapes experience, the Jacobian of Hypothesis~\ref{h:first}. Second, the chain rule gives exact decompositions that say where a change is felt and why. Third, feelings track internal inputs, not essential variables, and the two can disagree: feelings can mislead. Agreement must be supplied by learning or selection. When the gain is learned to sustain the body, damage comes to be felt in proportion to what it costs the whole (Experiment~\ref{exp:learned}), and, given a route from the reserve to the gain, learning uses it to suppress pain due to receptors as the reserve runs low (Experiment~\ref{exp:hunger}). When the gain learns to increase the scale, it makes injury feel good, and the body ultimately fails.

The account concerns valence alone. Arousal, the second dimension of core affect alongside valence~\citep{russell2003,barrett2017}, might speculatively relate to the spectral norm $\sigma_1(\bJ_\cS)$ of the Jacobian: responsiveness in the most responsive direction. Arousal narrows attention onto a few cues~\citep{easterbrook1959}, and, for a fixed scale $\|\bJ_\cS\|$, the spectral norm is largest when responsiveness concentrates along one direction and is smallest when it spreads evenly. Valence and arousal would then both be properties of $\bJ_\cS$.

Valence does not distinguish one feeling from another, fear from disgust, or dread from hunger (\S\ref{ssec:allo}). Accounts that construct emotions by interoceptive inference locate it in the inferred causes of interoceptive change~\citep{seth2013,barrett2017}; here it is left to the shape of $\bJ_\cS$, Hypothesis~\ref{h:first}, and to the variables through which changes arrive. The location of valence uses only part of the shape change. The rest changes the direction of the rows of $\bJ_\cS$, which parameters each action is sensitive to, and with it transparency and integrity, while leaving the scale and the shares of actions fixed. It says how interoception reshapes what the system can do, rather than how much or where, and is a candidate for what distinguishes one feeling from another. Open questions include whether and how such changes are felt in organisms. On the definitions here, learning only contributes to valence when it is driven by signals of what is at stake.

\paragraph{Which systems are self-sustaining?}
Reductions in power supply or physical damage move a language model toward its terminal set, but it is not self-sustaining: power is generated elsewhere and delivered along cables, nothing the model does affects that supply, and the model has no internal input that its supply, or damage to its activity, affects. Nor is its training felt: its loss is externally imposed, and no essential variable affects what drives its updates. Under Hypothesis~\ref{h:valence}, language models have no valence.

A battery-powered robot that recharges itself is not self-sustaining either: it runs at full responsiveness until its battery cuts out, so nothing drifts toward $Z$ until the cliff. A robot or drone whose signal of its charge scaled its responsiveness smoothly, and which had learned to recharge, would be self-sustaining. Whether it had valence would then turn on whether it is also a subject, which depends on how it is implemented~\citep{balduzzi2026g}. Valence is a matter of self-sustenance and subjectivity, not intelligence.

\paragraph{Measuring valence.}
Jacobians can be estimated by the size of the response to probes: evoked potentials, the amplitude of responses to transcranial magnetic stimulation~\citep{massimini2005,casali2013}, or pupil-linked measures of neural gain~\citep{astonjones2005, murphy2014, joshi2016}. When motion is a gain on activity, input and parameter Jacobians rescale together (Eq.~\eqref{eq:metabolism}), so the size of evoked responses tracks $\dot\lum$. The rate of change of the log of that size estimates the total motion of the scale, and its spatial pattern the split over locations; valence is the part due to internal inputs, which can be separated by blocking or driving a signal, as in Experiment~\ref{exp:reading}. The claims in Table~\ref{tab:pred} fail if the stated difference is absent or reversed.

\ifnum\arxivversion=0
\section*{Conflicts of interest}
\pending{to be added}

\section*{Funding}
\pending{to be added}

\fi

\section*{Code availability}
Code is available at \url{https://github.com/dbalduzzi/valence}.

\ifnum\arxivversion=0
\section*{Author contributions statement}
\pending{to be added}

\section*{Acknowledgements}
\pending{to be added}
\fi

\bibliographystyle{oup-abbrvnat}
\bibliography{refs}

\begin{appendices}
\section{Supplementary material}
\label{app:supp}

\subsection{Metabolism as a constraint}
\label{app:constraint}

Keep the energy loop, give the body two actions, one that costs the reserve and a forager that refills it, and let the network's parameters be learned, treating metabolism as a constraint. Suppose an organism has objectives $U_t$, set by reward signals and drives~\citep{keramati2014}, one of whose functions (hunger, say) is to act now so that energy does not reach dangerous levels later. It maximises reward over time subject to metabolic activity staying above $m_{\min}$:
\begin{gather}
  \label{eq:constraint}
  \max_{\cW}\ \sum_t U_t\big(\tilde{\by}_t\big)
  \quad\text{s.t.}\quad \\
  \phi_t(\cW) \coloneqq m_t(\cW) - m_{\min} \ge 0
  \quad \forall t . \notag
\end{gather}
$m_{\min}$ is not where death occurs; it is where it becomes necessary to \emph{do something}. The policy is trained by primal--dual updates, whose constraint gradient is $\nabla_\cW m_t$, weighted by multipliers $\mu_t$. The multipliers are homeostatic pressure: zero while energy is comfortable, and pressing harder the further it falls.

\suppexp{exp:constraint}{When the constraint binds} Only the costly action is rewarded, so the policy gains by spending the reserve, and must forage to keep $m$ above $m_{\min}$. Early on the reserve is full, the constraint slack, and valence close to zero. As the policy learns to spend, the reserve falls until the constraint binds, in every system; valence is then several times larger, and negative. The multipliers do not enter $v$; what changes is where the reserve sits. When it is full, $m$ saturates and barely moves (\S\ref{ssec:sign}); spent down, $m$ moves with every step.

\subsection{Both loops}
\label{app:learned}

Details of the body with both loops, \S\ref{ssec:learned}. Content is reduced to fixed rows, with norms $\bk$; valence depends only on the norms, so the rows can be any with integrity, such as those of Experiment~\ref{exp:pain}. Then $\|\bJ_\cS\|=m(E)\,\|\bh\odot\bG\odot\bk\|$; here, for simplicity, $m$ scales the Jacobian but not activity, so the reserve does not feed back on foraging. Activity at each location, $\tilde y_i=h_iG_i$, forages, refilling the reserve, and loads any injury, at the injured location and, through movement of the whole body, everywhere: $L_i=\tilde y_i+\bar{\tilde y}$. Repairing lost tissue draws on the reserve, $C_t\propto\sum_i(1-h_i)$. Receptors have no withdrawal ($\omega\equiv0$). Foraging yields $\bar{\tilde y}-\tfrac12\overline{\tilde y^2}$, which increases with activity only up to $1$. The gain parameters are bounded, $(a,b,c)=3\tanh\boldsymbol\theta$; the learned local part and the wirehead's global part settle at the bound.
\end{appendices}

\end{document}